\documentclass{article}

\usepackage{authblk}
\usepackage{amsmath}
\usepackage{amssymb}
\usepackage[dvips]{graphicx}
\usepackage{psfrag}
\usepackage{cite}
\usepackage{abstract}

 \newcommand{\Z}{\mathbf{Z}}
 \newcommand{\funcion}[3]{#1:\,#2\longrightarrow #3}
 \newcommand{\set}[2]{ \{\,#1\,|\,#2\,\}}
 \newcommand{\sset}[1]{ \{#1\} }
 \newcommand{\prima}{^\prime}
 \newcommand{\primas}{^{\prime\prime}}
 \newcommand{\nin}{ \not\in}
 \newcommand{\id}{\mathbf{1}}

 \newcommand{\ket}[1]{|#1\rangle}
 \newcommand{\bra}[1]{\langle #1|}
 \newcommand{\braket}[2]{\langle #1|#2 \rangle}

\newcommand{\pauli}[1]{G_{#1}}
\newcommand{\prob}[1]{\mathrm{Pr}(#1)}
\newcommand{\fr}{F_{\mathrm{r}}}
\newcommand{\fg}{F_{\mathrm{g}}}
\newcommand{\fb}{F_{\mathrm{b}}}
\newcommand{\pij}{{\langle i j\rangle}}
\newcommand{\tij}{{\langle i j k \rangle}}

\newcommand{\dual}[1]{{#1}^{\ast}}

\begin{document}

\date{}
\title{An Introduction to Topological Quantum Codes}
\author{H\'ector Bomb\'in}
\affil{Perimeter Institute for Theoretical Physics\\ 31 Caroline St. N., Waterloo, ON, N2L 2Y5, Canada}

\maketitle

\begin{abstract}
This is the chapter \emph{Topological Codes} of the book \emph{Quantum Error Correction}, edited by Daniel A. Lidar and Todd A. Brun, Cambridge University Press, New York, 2013.
\\
\\
http://www.cambridge.org/us/academic/subjects/physics/quantum-physics-
quantum-information-and-quantum-computation/quantum-error-correction
\\
\\
\copyright \,Cambridge University Press
\\
This publication is in copyright. Subject to statutory exception and to the provisions of relevant collective licensing agreements, no reproduction of any part may take place without the written permission of Cambridge University Press.
\end{abstract}

\newpage
\tableofcontents

\newpage

\section{Introduction}

What a good code is depends on the particular constraints of the problem at hand. In this chapter we address a constraint that is relevant to many physical settings: locality. In particular, we are interested in situations where  \emph{geometrical locality} is relevant. This typically means that the physical qubits composing the code are placed in some lattice and only interactions between nearby qubits are possible. In this case, it is desirable that syndrome extraction also be local, so that fault tolerance can possibly be achieved. Topological codes offer a natural solution to locality constraints, as they have stabilizer generators with local support.

\begin{figure*}
\center
\psfrag{a}{\normalsize $a$}
\psfrag{b}{\normalsize $b$}
 \includegraphics[width=7cm]{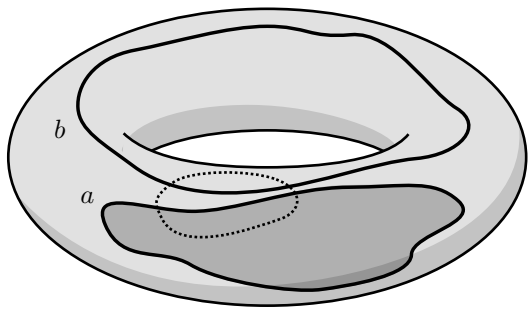}
\caption{Closed curves in a torus can be the boundary of a region, like curve $a$. However, curve $b$ is not the boundary of a region. The difference cannot be detected by looking only at a local region such as the one marked with a dotted line.}
\label{fig:idea}
\end{figure*}

In topological codes information is stored in \emph{global} degrees of freedom, so larger lattices provide larger code distances. The nature of these global degrees of freedom is illustrated in Fig.~\ref{fig:idea}, where several closed curves in a torus are compared. Consider curves $a$ and $b$. They look the same if examined locally, as in the region marked with dotted lines. However, curve $a$ is the \emph{boundary} of a region, whereas curve $b$ is not. In order to decide whether a curve is a boundary or not we need global information about it. This is, as we will see, a core idea in topological codes.

This chapter only attempts to provide an introduction to the subject. In particular, we will only deal with two-dimensional codes and leave out the condensed-matter perspective.

\section{Local Codes}

Since the emphasis of this chapter is on locality, we start by giving a formal definition of what it means for a code to be local. Intuitively, a local stabilizer code is one in which all stabilizer generators act only on a few nearby qubits. To formalize this idea, we could talk about $n$-local codes, those in which the support of each stabilizer generator is limited to $n$ qubits. This is a notion that can be applied to individual codes. But if we want to talk about local codes, without further adjectives, we have to consider instead sets or \emph{families} of codes.

\begin{figure*}
\center
 \includegraphics[width=5.5cm]{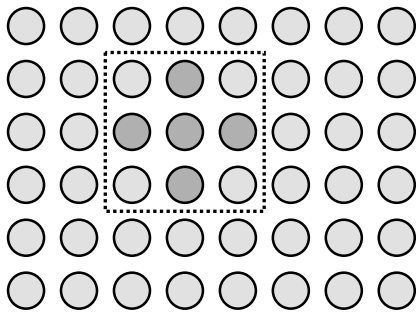}
\caption{In 2D local codes qubits are arranged in a 2D array. The support of any stabilizer generators must be contained in a box of a fixed size, here a $3\times 3$ box. Circles represent qubits and darker ones form the support of a generator. }
\label{fig:locality}
\end{figure*}

A family of stabilizer codes is \emph{local} if we can choose the stabilizer generators so that:
\begin{enumerate}
\item  the family contains codes of arbitrary large distance,
\item the number of qubits in the support of the stabilizer generators is bounded, and
\item the number of stabilizers  with support containing any given qubit is bounded.
\end{enumerate}
This notion of locality can be put in terms of graph connectivity, without any further structure. But in practice we are often interested in locality from a geometrical point of view. A family of codes is \emph{local in $D$ dimensions} when
\begin{enumerate}
\item qubits are placed in a D-dimensional array and
\item the support of any stabilizer generator is contained in a hypercube of bounded size.
\end{enumerate}
This is illustrated in Fig.~\ref{fig:locality}. Notice that a code that is local in a geometric sense is also local in the more general sense above.

Locality does not come without a price. For example, in any family of 2D local codes the distance is $d=O(\sqrt n)$, with $n$ the total number of qubits. We do not prove this here, but see \cite{Bravyi:2009:043029}. This behavior of the code distance might appear undesirable, but indeed it is not harmful because the code distance alone does not dictate the error correcting capability of a family of topological codes.  The key for fault tolerance is statistics: an error that cannot be corrected but is unlikely to occur is not important. As we will see, topological codes can correct \emph{most} errors of weight $O(n)$, which is reflected in the existence of an error threshold. For noise below the threshold, error correction can be achieved with asymptotically perfect accuracy in the limit of large lattices.

\section{Surface Homology}\label{sec:homology}

In order to understand topological codes, it is convenient to have some background in algebraic topology. This section provides an elementary introduction to the topic, which will be sufficient for our purpose. For further reading, see for example \cite{Hatcher:2002:CUP}.

\subsection{Topology of Closed Surfaces}

Topology deals with spatial properties that are preserved under continuous deformations. It is sometimes called ``rubber sheet geometry'', because we are allowed to stretch or compress our objects of study, but not to tear them or glue their parts. The coffee mug and the donut are well known examples that look the same to a topologist since, up to deformations, they are both nothing but a solid sphere with a hole.

In the topological codes that we shall study qubits are placed on two-dimensional lattices. Such lattices will be embedded in surfaces, and it turns out that the topology of the surfaces is what matters to us. Since we only have a finite number of qubits at hand, there is no point in considering open surfaces, like the plane. Thus, we restrict ourselves to closed surfaces, like the sphere. For simplicity, we will focus on orientable closed surfaces here. These are the closed surfaces that have an inside and an outside, that is, those that are the boundaries of everyday solid objects. Moreover, we only consider connected surfaces, those in which we can move from any point to another without jumps.

\begin{figure*}
\center
 \includegraphics[width=9cm]{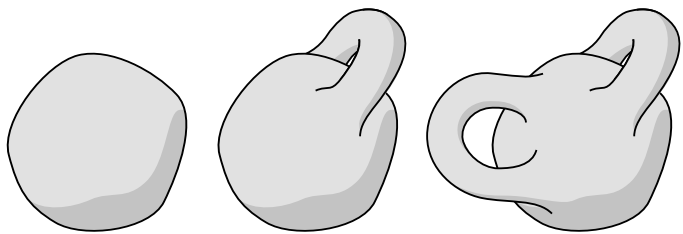}
\caption{From the topological perspective, orientable closed surfaces are classified by the genus $g$ or number of handles. These are the three with lower genus: the sphere, the sphere with a handle or torus, and the sphere with two handles or 2-torus.}
\label{fig:surfaces}
\end{figure*}

From a topological perspective, the classification of connected orientable closed surfaces is pretty simple. First, we have the sphere. If we add a handle to a sphere, we obtain the torus (the surface of a donut). We can then add a second handle (2-torus), or a third (3-torus), and so on, see Fig.~\ref{fig:surfaces}. This infinite process allows us to build all orientable closed surfaces, which are thus classified by the number of handles, known as their \emph{genus}. A sphere has genus 0 and a $g$-torus has genus $g$. As we will see, the genus of a surface will dictate the number of encoded qubits in a topological code.

There is an interesting relationship between the number of elements of a lattice embedded in a surface and its genus. We will denote the number of vertices, edges and faces of the lattice as $V$, $E$ and $F$, respectively. The Euler characteristic is then defined as the quantity
\begin{equation}\label{Euler}
\chi:=V-E+F.
\end{equation}
This is an important quantity because it only depends on the topology of the surface, not on the particular lattice. In particular, for closed orientable surfaces we have
\begin{equation}\label{Euler_genus}
\chi=2(1-g).
\end{equation}

\subsection{Homology of Curves}

Suppose that you are given two objects and asked whether they are topologically equivalent. If you can find a way to continuously transform one into the other, you will have shown that they are equivalent. But, if they are not equivalent, how can we show it? Topological  \emph{invariants} offer an aswer to this question. A topological invariant is a number or other kind of mathematical structure that is constructed from the objects of interest and depends only on their topology. If two objects return different values for any topological invariant, they cannot be topologically equivalent. Notice that we have already encountered an example of topological invariant, namely, the Euler characteristic.

The topological invariant that is involved in the construction of the most basic topological codes is the \emph{first homology group} of a surface. Actually, the elements of this group label a basis for the encoded states, as we will see. So, what is homology about?

\begin{figure*}
\center
\psfrag{a}{\normalsize $a$}
\psfrag{b}{\normalsize $b$}
\psfrag{c}{\normalsize $c$}
\psfrag{d}{\normalsize $d$}
\psfrag{A}{\normalsize $A$}
\psfrag{B}{\normalsize $B$}
 \includegraphics[width=7cm]{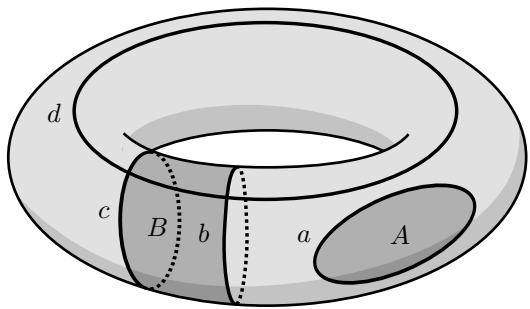}
\caption{Several closed curves in a torus. Curve $a$ is the boundary of region A, so it is homologically trivial. Curves $b$ and $c$ form the boundary of region $B$, and thus are homologically equivalent. Curve $d$ is homologically nontrivial and not equivalent to $c$.}
\label{fig:homology}
\end{figure*}

Consider the closed curves on a torus of Fig.~\ref{fig:homology}. Curve $a$ is the \emph{boundary} of region $A$, whereas curves $b, c, d$ are not the boundaries of any region (individually). We say that $a$ is homologically trivial, and $b,c,d$ homologically nontrivial. Moreover, $b$ and $c$ together enclose the region $B$, so they are said to be homologically equivalent. On the other hand, $c$ and $d$ together do not form the boundary of any region and are thereof not equivalent. We have thus a classification of the closed curves on a surface into different \emph{homology classes}. We can further give to these classes the structure of an abelian group, with the identity element corresponding to the trivial homology class of curves. This requires adopting a more formal approach.

We will work now with a particular lattice embedded in the surface. As it is customary, we rename vertices as 0-cells, edges as 1-cells and faces as 2-cells. First, we want to form an abelian group $C_i$ out of the set of $i$-cells, for $i=0,1,2$. Consider for example 1-cells. If we label the edges as $\sset{e_i}_1^E$ we can represent any set of edges $E\prima$ as a formal sum
\begin{equation}\label{1chain}
c=\sum_i c_i \,e_i,\qquad\qquad c_i =  \left\{
\begin{array}{c l}
  0 & e_i\nin E\prima \\
  1 & e_i\in E\prima
\end{array}
\right.
.
\end{equation}
Such formal sums are called 1-chains. They can be added together to obtain another 1-chain taking into account the rule $e_i+e_i=0$. This gives rise to an abelian group structure on the 1-chains $C_1$, with the zero element corresponding to the empty set. We can represent 1-chains as binary vectors of length $E$, and indeed $C_1\simeq\Z_2^E$, where $\Z_2$ is $F_2$ considered just as an additive group. A nice aspect of the notation \eqref{1chain} is that any edge $e$ automatically denotes the $1$-chain corresponding to the set $\sset{e}$. Similarly, we can form the group of $0$-chains $C_0\simeq \Z_2^V$ and the group of $2$-chains $C_2\simeq \Z_2^F$.

\begin{figure*}
\center
\psfrag{d1}{\normalsize $\partial_1$}
\psfrag{d2}{\normalsize $\partial_2$}
 \includegraphics[width=8cm]{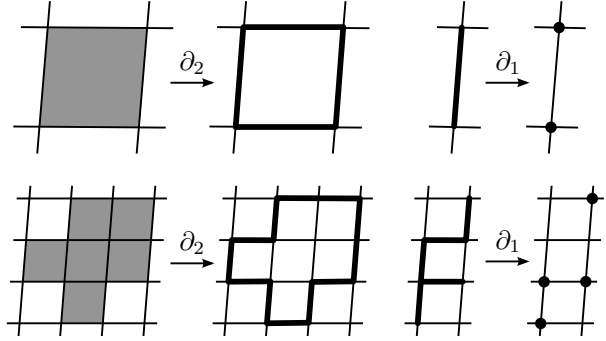}
 \caption{The action of boundary operators. $\partial_2$ maps a set of faces to the set of edges that form its boundary. $\partial_1$ maps a set of edges to to the set of vertices where an odd number of these edges meet.}
\label{fig:boundary}
\end{figure*}

Our next step is to introduce a family of group homomorphisms $\partial$, called boundary operators. There are actually two that are relevant to our discussion,
\begin{equation}
\funcion {\partial_2}{C_2}{C_1},\qquad\qquad \funcion {\partial_1}{C_1}{C_0},
\end{equation}
but they are commonly referred as $\partial$. As its name suggests, $\partial$ takes objects to their boundaries, as illustrated in Fig.~\ref{fig:boundary}. Namely, if a face $f$ has as boundary the set of edges $\sset{e\prima_1,\dots,e\prima_k}$, then $\partial_2 f=e\prima_1+\dots+e\prima_k$. What happens when we apply $\partial_2$ to a region, that is, to a set of faces $F\prima=\sset{f\prima_1,\dots,f\prima_l}$? As a 2-chain, the region has the expression $r=f\prima_1+\dots+f\prima_l$, so $\partial_2r=\partial f\prima_1+\dots+\partial f\prima_l$ because $\partial$ respects the abelian group structure. Since $e_i+e_i=0$, the edges that are shared by neighboring faces in $F\prima$ cancel and we have $\partial_2 r=e\primas_1+\dots+e\primas_m$, where $\sset{e\primas_1,\dots,e\primas_m}$ is the set of edges that form the boundary of the region. The definition of $\partial_1$ is analogous: if the edge $e$ has as endpoints the vertices $v\prima_1, v\prima_2$, then $\partial_1 e= v\prima_1+v\prima_2$. For sets of edges the boundary is composed of those vertices at which an odd number of these edges meet.

Now we define two subgroups of $1$-chains $Z_1,B_1\subset C_1$. $Z_1$ is the subgroup of $1$-chains $z$ that have no boundary, that is, $\partial_1 z=0$, the kernel of $\partial_1$. The elements of $Z_1$, called cycles, are collections of closed curves. $B_1$ is the image of $\partial_2$, that is, the subgroup of $1$-chains $b$ that are a boundary of a $2$-chain, $b=\partial_2 c$ for some $c\in C_2$. The crucial observation is that all boundaries are also cycles, namely $B_1\subset Z_1$. In other words,
\begin{equation}
\partial^2:=\partial_1\circ\partial_2=0.
\end{equation}
We can then define the first homology group $H_1$ as the quotient
\begin{equation}\label{eq:homology}
H_1:= Z_1/B_1.
\end{equation}
 Let us recall the quotient group construction for abelian groups. The elements of $H_1$ are cosets of the form $\overline z:=\set{z+b}{b\in B_1}$ for some $z\in Z_1$, the addition is that inherited from $Z_1$, namely $\overline z+\overline z\prima = \overline{z+z\prima}$, and the zero element is the coset $\overline 0=B_1$. Technically, \eqref{eq:homology} defines $H_1(S;\Z_2)$, the first homology group of the surface $S$ over $\Z_2$.

Algebraic topology teaches us that the group $H_1$ only depends, up to isomorphisms, on the topology of the surface. Indeed:
\begin{equation}\label{eq:homology_g}
H_1 \simeq \Z_2^{2g}.
\end{equation}
For a sphere, $g=0$ and thus the first homology group is trivial: all cycles are boundaries, $B_1=Z_1$. Notice how taking the quotient has erased all the information about the particular lattice used: this is the magic of topological invariants.

\begin{figure*}
\center
\psfrag{a}{\normalsize $a$}
\psfrag{b}{\normalsize $b$}
\psfrag{c}{\normalsize $c$}
\psfrag{d}{\normalsize $d$}
\psfrag{e}{\normalsize $e$}
\psfrag{f}{\normalsize $f$}
\psfrag{A}{\normalsize $A$}
\psfrag{B}{\normalsize $B$}
\psfrag{C}{\normalsize $C$}
 \includegraphics[width=7cm]{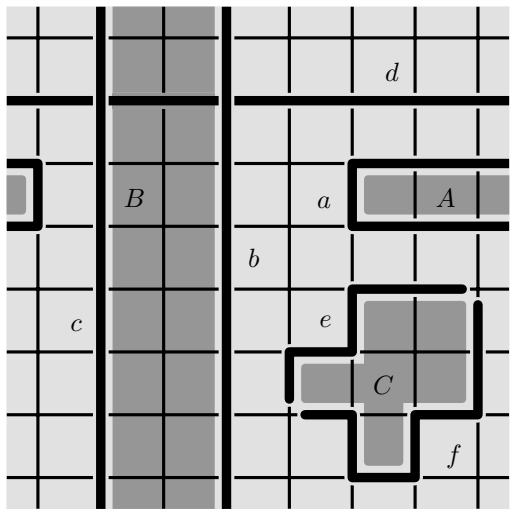}
\caption{Several 1-chains in a periodic $8\times 8$ square lattice. $\overline a=0$ because $a=\partial A$. $\overline b=\overline c$ because $b+c=\partial B$. $\overline c\neq \overline d\neq 0$ because neither $c+d=\partial D$, $c=\partial D$ nor $c=\partial D$ for any region $D$. $\overline e=\overline f$ because $e+f=\partial C$.}
\label{fig:homology_lattice}
\end{figure*}

As an example, consider the square lattice embedded in a torus in Fig.~\ref{fig:homology_lattice}. Here we are representing the torus in a convenient and conventional way, as a square where opposite edges are identified. The connected 1-chains in the figure are subject to the same homological equivalences as in Fig.~\ref{fig:homology}. In the notation we have just developed, we have $\overline a=0$, $\overline b = \overline c$ and $\overline b\neq \overline d\neq 0$.  For a torus $H_1\simeq \Z_2\times \Z_2$, so the homology group has two generators. Here we can take, for example, $\overline b$ and $\overline d$ as the generators, and the the elements of $H_1$ are $0,\overline b,\overline d$ and $\overline b+\overline d$.

The notion of equivalence up to homology is also useful for 1-chains that are not cycles. That is, we can consider the quotient $C_1/B_1$, and use the notation $\overline c := \set{c+b}{b\in B_1}$, where $c\in C_1$, to denote its elements. The open 1-chains $e$ and $f$ in Fig.~\ref{fig:homology_lattice} are homologous: they enclose the region $C$.

We have already encountered a quotient construction before, when studying stabilizer codes in Chapter 2, where logical operators are recovered from the quotient $N(S)/S$. As we will see in the next section, this quotient construction and the one in \eqref{eq:homology} can be fruitfully related.

\section{Surface Codes}\label{sec:surface_codes}

Surface codes are the most basic examples of topological codes.  In this section we will motivate their construction and study their main features.

\subsection{Definition}

The idea behind surface codes is to encode information in ``homological degrees of freedom''. To this end, our first step is to fix a lattice embedded in a given closed surface. We attach a qubit to each edge, so each element of the computational basis can be interpreted as a 1-chain $c\in C_1$ in the most obvious manner:
\begin{equation}\label{1chain_basis}
\ket {c}:= \bigotimes_i \ket {c_i},\qquad\qquad c\in C_1.
\end{equation}
Here $i$ runs over the physical qubits or, equivalently, the edges of the lattice $\sset{e_i}$.
We will find it convenient to label products of $X$ and $Z$ Pauli operators with 1-chains, too:
\begin{equation}\label{XZ_ops_1chain}
X_c:= \bigotimes_i X_i^{c_i} ,\qquad Z_c:= \bigotimes_i Z_i^{c_i} ,\qquad\qquad c\in C_1.
\end{equation}
Notice that these labelings are indeed group homomorphisms from $C_1$ to $\pauli n$, because
\begin{equation}\label{XZ_ops_1chain2}
X_cX_{c\prima}=X_{c+c\prima},\qquad\qquad Z_cZ_{c\prima}=Z_{c+c\prima}.
\end{equation}
The definition of the surface code is quite natural. It has a basis with elements (here and anywhere else we ignore normalization)
\begin{equation}\label{surface_code_basis}
\ket {\overline z}:= \sum_{b\in B_1}\,\ket {z+b}, \qquad\qquad \overline z\in H_1,
\end{equation}
which are sums of all the cycles that form a given homology class. Clearly $\braket {\overline z}{\overline z\prima}=0$ for $\overline z\neq \overline z\prima$. Then from \eqref{eq:homology_g} we have $|H_1|=2^{2g}$ and  the number of encoded qubits is $k=2g$.

To get a first flavor of the power of surface codes, we will study the effect of bit-flip errors. Notice that $X_c\ket b= \ket{b+c}$ and thus $X_c\ket{\overline z}= \ket{\overline z+\overline c}$. Let $z,z\prima \in Z_1$. If $\partial c\neq 0$ then $\partial(z+c)\neq 0$ and $\overline z+ \overline c\neq \overline z\prima$. This implies $\bra {\overline z\prima} X_c \ket {\overline z}=0$, so $X_c$ can map the code to itself only if $c$ is a cycle. But errors $X_b$ with $b$ a boundary do nothing, as $X_b\ket{\overline z}=\ket {\overline z+\overline b}=\ket {\overline z}$. Therefore, only bit-flip errors $X_z$ with $z\in Z_1$ are undetectable. It follows that the distance of a surface code for bit-flip errors is the length of the shortest nontrivial cycle. A similar analysis can be done for $Z$ errors, but we will postpone the discussion until we can use the language of stabilizer operators. We anticipate that the distance for $Z$ errors is given by the shortest nontrivial cycle in the \emph{dual} lattice.

\subsection{Stabilizer Group}

\begin{figure*}
\center
\psfrag{v}{\normalsize $v$}
\psfrag{f}{\normalsize $f$}
\psfrag{Zv}{\normalsize $Z_v$}
\psfrag{Xf}{\normalsize $X_f$}
 \includegraphics[width=6cm]{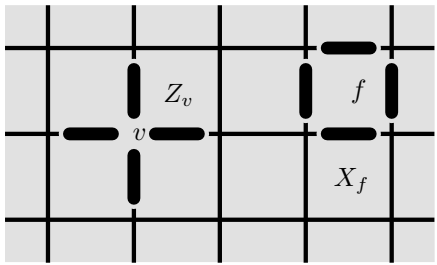}
\caption{The support of vertex and face stabilizer generators.}
\label{fig:star_plaquette}
\end{figure*}

Given a vertex $v$ and a face $f$, consider the face (``plaquette'') and vertex (``star'') Pauli operators
\begin{equation}\label{surface_code_stabilizers}
X_f:= \prod_{e\in \partial_2 f}\,X_e, \qquad\qquad Z_v:= \prod_{e | v\in \partial_1 e}\,Z_e,
\end{equation}
where $\partial f$ and $\partial e$ are understood as sets. These operators are depicted in Fig.~\ref{fig:star_plaquette}, where we can see that a vertex operator has support on the links that meet at a vertex and a face operator has support on the edges that enclose the face. Using the notation \eqref{XZ_ops_1chain}, we could have written $X_f:= X_{\partial f}$ in \eqref{surface_code_stabilizers}, but we wanted to remark on the symmetry between vertex and face operators.

Vertex and face operators commute with each other. We claim that they generate the stabilizer of surface codes, as defined in \eqref{surface_code_basis}. To check this explicitly,  consider the states
\begin{equation}\label{surface_code_projection}
\ket {\tilde c}:=  \prod_{f}\frac {1+X_f}2 \,\prod_{v}\frac {1+Z_v}2\,\ket c,\qquad\qquad c\in C_1.
\end{equation}
These states span the code defined by the stabilizers \eqref{surface_code_stabilizers}, because $\ket {\tilde c}$ is the projection of $\ket c$ onto the code subspace. Notice that $Z_v\ket c=-\ket c$ if $v\in \partial c$, and $Z_v\ket c= \ket c$ otherwise. Therefore, $\ket {\tilde c}=0$ if $\partial c\neq 0$. We next notice that the first product in \eqref{surface_code_projection} can be expanded as a sum over subsets of faces $\sset{f_i}$:
\begin{equation}\label{surface_code_projection2}
\prod_{f} (1+X_f) =  \sum_{\sset{f_i}} \prod_i X_{\partial_2 f_i} =   \sum_{\sset{f_i}} X_{\partial_2 (\sum_i f_i)} = \sum_{c_2 \in C_2} X_{\partial_2 c_2},
\end{equation}
where we have used \eqref{XZ_ops_1chain2}. Since $\partial_2$ is a group homomorphism, we can replace the sum over 2-chains by a sum over 1-chains in the image of $\partial_2$, up to a factor. Therefore, for $z\in Z_1$ we have
\begin{equation}\label{surface_code_projection3}
\ket {\tilde z}:=  \prod_{f}\frac {(1+X_f)}2 \,\ket z\propto\sum_{b\in B_1} X_b \,\ket z=\ket {\overline z}.
\end{equation}
We get the same code subspace from the stabilizers \eqref{surface_code_stabilizers} and the span of the basis \eqref{surface_code_basis}.

Notice the different role played by vertex and face operators. Face operators are related to $\partial_2$, and they stabilize the subspace with basis $\ket {\overline c}:=\sum_{b\in B_1} \ket {c+b}$, $c\in C_1$. That is, face operators enforce that states should be a uniform superposition of states on the same homology class. Vertex operators are related to $\partial_1$, and they stabilize the subspace with basis $\ket z$, $z\in Z_1$. That is, vertex operators enforce that states should have no boundary.

The stabilizer generators \eqref{surface_code_stabilizers} are not all independent. As one can easily check, they are subject to the following two conditions, and no more:
\begin{equation}\label{surface_code_stabilizers_constraint}
 \prod_f \, X_f=1,\qquad\qquad \prod_v\,Z_v=1.
\end{equation}
Therefore, there are $V+F-2$ independent generators. It follows from the theory of stabilizer codes that the number of encoded qubits is
\begin{equation}
k=E-(V+F-2)=2-\chi = 2g,
\end{equation}
which of course agrees with the value given after \eqref{surface_code_basis}.

\subsection{Dual lattice}

\begin{figure*}
\center
 \includegraphics[width=6cm]{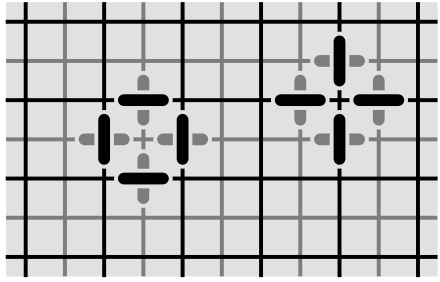}
\caption{A lattice and its dual. Under duality, vertex and faces are interchanged, and so are vertex and face operators.}
\label{fig:dual_lattice}
\end{figure*}

Given a lattice embedded in a surface, we can construct its dual lattice. This is illustrated in Fig.~\ref{fig:dual_lattice}. The idea is that the faces of the original lattice get mapped to vertices in the dual lattice, edges to dual edges, and vertices to dual faces. We will use a hat $\dual  {}$ to denote dual vertices $\dual f$, dual edges $\dual e$ and dual faces $\dual v$, in terms of their related faces $f$, edges $e$ and vertices $v$ of the original lattice, respectively. Similarly, we have dual boundary operators
\begin{equation}
\funcion {\dual \partial_1}{\dual C_0}{\dual C_1},\qquad\qquad\funcion {\dual \partial_2}{\dual C_1}{\dual C_2}
\end{equation}
acting on dual chains $\dual c$. \emph{To simplify notation, for generic 1-chains we will consider $\dual c$ and $c$ to be unrelated objects}. But, for single edges, $\dual e$ denotes the dual of $e$, so $\dual e$ and $e$ refer to the same physical qubit. The boundary operators $\dual \partial$ produce the groups of dual cycles $\dual Z_1$ and dual boundaries $\dual B_1$, and thus a homology group
\begin{equation}
\dual H_1=\frac {\dual Z_1}{\dual B_1} \simeq H_1.
\end{equation}

Comparing the action of $\partial$ and $\dual\partial$ on their respective lattices, we observe that
\begin{equation}\label{dual_boundary}
\dual e\in \dual \partial_1 \dual v \iff v\in \partial_1 e,\qquad\qquad \dual f\in \dual \partial_2 \dual e \iff e\in \partial_2 f.
\end{equation}
Now, consider the effect of applying a transversal Hadamard gate $W^{\otimes E}$ across all qubits in a surface code. The code is mapped to a new subspace, described by the stabilizers:
\begin{alignat}{3}\label{surface_code_stabilizers_W}
W^{\otimes E} X_f W^{\otimes E} &=  \prod_{e\in \partial_2 f}\,Z_e &= \prod_{\dual e| \dual f\in \dual \partial_2 \dual e} Z_e &=: Z_{\dual f}\nonumber\\
W^{\otimes E} Z_v W^{\otimes E} &=  \prod_{e| v\in \partial_1 e}\,X_e &= \prod_{\dual e \in \dual \partial_1 \dual v} X_e &=: X_{\dual v}.
\end{alignat}
This is nothing but the surface code defined on the dual lattice! The moral is that we can deal with phase-flip errors as we already did with $X$ errors, but working in the dual lattice. As a result, we have that the distance of a surface code is the length of the shortest nontrivial cycle in the original or the dual lattice.

As an example, consider periodic square lattices of size $d\times d$ embedded in a torus, like the one in Fig.~\ref{fig:toric_code}. These lattices produce what was the first example of surface codes, the family of ``toric codes''. The dual of the $d\times d$ square lattice is again a $d\times d$ square lattice, and thus the distance is $d$ because nontrivial cycles have to wind around the torus, as shown in the figure. It follows that these $[[2d^2,2,d]]$ codes form a family of 2D local codes.

\subsection{Logical operators}

In the previous section we have learned that while it is useful to relate bit-flip errors $X_c$ to 1-chains $c$, phase-flip errors should be related to 1-chains $\dual c$ in the dual lattice. Thus, we will use the notation $Z_{\dual c}:=\prod_{i} Z_{e_i}$, where $\dual c=\sum_i {\dual e_i}$ for some edge subset $\sset{e_i}$. Any Pauli operator $A$ can be written as
\begin{equation}\label{Pauli}
A=i^{\alpha} X_{c} Z_{\dual c},\qquad\qquad (c,\dual c) \in C_1\times\dual C_1,\quad\alpha\in Z_4.
\end{equation}
Therefore, any Pauli operator can be visualized as a pair of chains $(c,\dual c)$ or, to give it a physical flavor, as a collection of strings. These strings are of two kinds, as they can live in the direct lattice (the $X$ part) or the dual lattice (the $Z$ part). Strings can be open, if they have endpoints, or closed, if they form a loop. An important property of closed string operators is that a direct and a dual string anticommute if and only if they cross an odd number of times. Notice that the oddness of the number of crossings is preserved up to homology. This has to be the case because, for $(b,\dual b) \in B_1\times\dual B_1$ and $(z,\dual z) \in Z_1\times\dual Z_1$, $X_z$ and $Z_{\dual z}$ commute if and only if $X_{z+b}$ and $Z_{\dual z+\dual b}$ commute.

\begin{figure*}
\center
\psfrag{Z1}{\normalsize $\overline Z_1$}
\psfrag{X1}{\normalsize $\overline X_1$}
\psfrag{Z2}{\normalsize $\overline Z_2$}
\psfrag{X2}{\normalsize $\overline X_2$}
 \includegraphics[width=7cm]{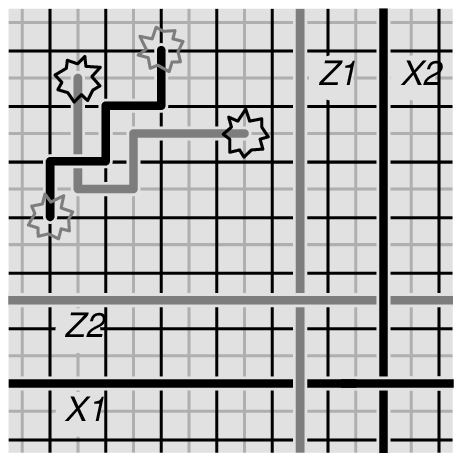}
\caption{String operators in a toric code. $X$-type ($Z$-type) string operators belong to the direct (dual) lattice and are displayed in a darker (softer) tone. The two open string operators anticommute with the vertex or face operators at their endpoints, marked with stars. Logical operators take the form of nontrivial closed string operators. The labeling agrees with the fact that crossing string operators of different types anticommute. This is a $[[128, 2, 8]]$ code.}
\label{fig:toric_code}
\end{figure*}

The relationship between $N(S)/S$ and $Z_1/B_1$ will now become clear. First, it is easy to check that
\begin{equation}\label{XZ_ops_boundary}
[X_c,Z_v]=0\iff v\nin \partial_1 c,\qquad [Z_{\dual c},X_{\dual f}]=0\iff \dual f\nin \dual \partial_2 \dual c,
\end{equation}
as illustrated in Fig.~\ref{fig:toric_code}. Consider any Pauli operator $A$ as in \eqref{Pauli}. It follows from \eqref{XZ_ops_boundary} that $A\in N(S)$ if and only if $(c,\dual c) \in Z_1\times \dual Z_1$. What about the elements of the stabilizer $S$? An arbitrary stabilizer element will have the form $B=\prod_{i} X_{f_i}\prod_{j} Z_{v_j}$ for some subset of faces $\sset{f_i}$ and some subset of vertices $\sset{v_j}$. We can apply the same trick as in \eqref{surface_code_projection2}, obtaining $B=X_{\partial_2 c_2}Z_{\dual \partial_1 \dual c_0}$, where $c_2=\sum_i f_i$ and $\dual c_0=\sum_i \dual v_i$. Therefore, we see that $A$ belongs to $S$ if and only if $\alpha=0$ and $(c,\dual c)\in B_1\times \dual B_1$.

In summary, we have just seen that the elements of the normalizer $N(S)$ are labeled, up to a phase, by a cycle and a dual cycle, whereas the elements of the stabilizer are labeled by a boundary and a dual boundary. The parallelism is now apparent: we can label the elements of $N(S)/S$, up to a phase, with an element of $H_1\times\dual H_1$. The cosets indeed take the form
\begin{equation}\label{}
\set {i^{\alpha}X_{z+b}Z_{\dual z+\dual b}}{(b,\dual b)\in B_1\times\dual B_1, \alpha \in \Z_4},\qquad (\overline z,\overline{\dual z})\in H_1\times \dual H_1.
\end{equation}
Setting $S\prima:=\langle i \id \rangle S$, so that $S\prima$ is the center of $N(S)$, we have
\begin{equation}\label{}
\frac{N(S)}{S\prima}\simeq H_1\times\dual H_1 \simeq H_1^2.
\end{equation}

Recall from the theory of stabilizer codes that $N(S)/S$ describes logical Pauli operators. In surface codes, we may therefore choose as generators of logical Pauli operators a set of closed string operators, as in the toric code of Fig.~\ref{fig:toric_code}. 
Notice how the crossings and the required commutation relations match.

\subsection{Boundaries}\label{sec:boundaries}

From a practical perspective, the 2D locality of surface codes can be very useful. However, the fact that we need nontrivial topologies complicates things from a geometrical perspective: it might be difficult to place qubits in a toroidal geometry. Fortunately, this obstacle can be overcome by introducing boundaries, as we will see. Boundaries can create nontrivial topologies even in a plane, which makes it possible to construct planar families of surface codes. Since a homological description of boundaries requires discussing the concept of relative homology, we will take an alternative route based on string operators.

\begin{figure*}
\center
\psfrag{Z}{\normalsize $\overline Z$}
\psfrag{X}{\normalsize $\overline X$}
\psfrag{f}{\normalsize $f$}
\psfrag{f2}{\normalsize $f\prima$}
 \includegraphics[width=7cm]{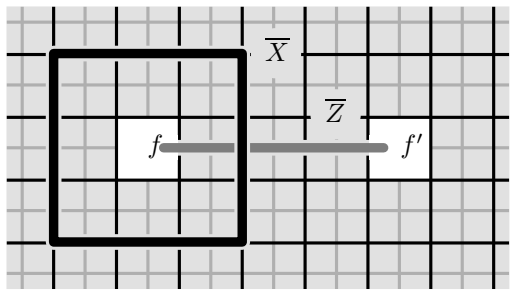}
\caption{When we remove two face stabilizer generators, a new encoded qubit appears. A logical operator takes the form of an open dual string connecting the missing faces and a direct closed string enclosing one of the holes.}
\label{fig:missing_generators}
\end{figure*}

In order to motivate the construction, we start with the following simple example. Suppose that, in a surface code, we remove the stabilizer generators corresponding to two separate faces $f$ and $f\prima$. We know that this must increase the number of encoded qubits by one because, taking \eqref{surface_code_stabilizers_constraint} into account, there is one generator less. Which are the string operators for this encoded qubit? The answer can be found in Fig.~\ref {fig:missing_generators}: a dual string operator with endpoints in the faces belongs now to $N(S)$. And a direct string that encloses $f$ no longer belongs to $S$, since it is the product of either all the face operators ``inside'' it, which include $X_{f}$, or all the face operators ``outside'' it, which include $X_{f\prima}$. These two strings cross once and thus provide the $\overline X$ and $\overline Z$ operators for the new encoded qubit.

\begin{figure*}
\center
\psfrag{X1}{\normalsize $X_1$}
\psfrag{X2}{\normalsize $X_2$}
\psfrag{Z1}{\normalsize $Z_1$}
 \includegraphics[width=7cm]{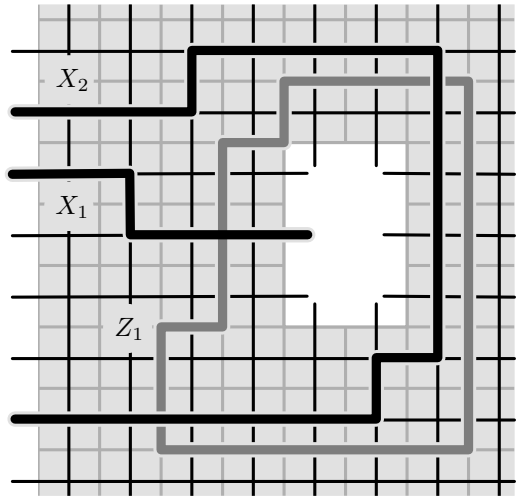}
\caption{A piece of a surface code with two dual or ``rough'' boundaries, one of them in the form of a hole. The direct string operator $X_1$ belongs to $N(S)$ because its endpoints lie on the dual boundaries, but does not belong to $S$ as it does not enclose any region. The dual string operator $Z_1$ is closed, so $Z_1\in N(S)$. It encloses a region, but this contains a piece of dual boundary and thus $Z_1\nin S$. Indeed, $\sset {X_1,Z_1}=0$ because the strings cross. As for the direct string operator $X_2$, it encloses a region that only contains dual boundaries, so $X_2\in S$.}
\label{fig:rough}
\end{figure*}

What is the distance of the new code? As we separate the two faces, the distance for phase-flip errors will grow accordingly. However, bit-flips do not behave like that, because a string that encloses $f$ can be very small. Indeed, the smallest possible such string operator is $X_{f}$ itself. Thus, we have failed at introducing an entirely global degree of freedom. Fortunately, the solution is at hand. If the perimeter of the faces $f$ and $f\prima$ is large,  the distance for phase-flip errors will be large too.

The lesson is that we can introduce a nontrivial topology in the lattice by ``erasing big faces''. If we start with a sphere and remove $h+1$ faces we will end up with a disc with $h$ holes. The resulting surface code encodes $h$ qubits. Naturally, we can do the same constructions in the dual lattice: removing $r+1$ vertices will introduce $r$ new encoded qubits. Because of their appearance, boundaries in the dual lattice are sometimes called ``rough'', whereas boundaries in the direct lattice are called ``soft''. Fig.~\ref{fig:rough} shows an example of a geometry with dual boundaries.

\begin{figure*}
\center
\psfrag{X1}{\normalsize $\overline X$}
\psfrag{Z1}{\normalsize $\overline Z$}
 \includegraphics[width=7cm]{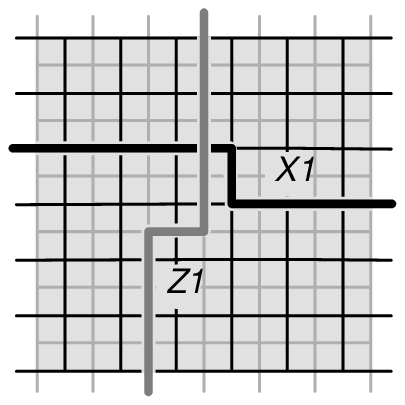}
\caption{A planar toric code. The top and bottom boundaries are direct, the left and right dual. Nontrivial direct (dual) strings connect left and right (top and bottom) boundaries, so that this is a $[[85,1,7]]$ code.}
\label{fig:planar_toric_code}
\end{figure*}

It is possible to describe boundaries in term of how they modify the notion of closed and boundary strings. For example, direct (soft) boundaries give rise to the following properties:
\begin{enumerate}
\item Dual string operators that have their two endpoints on a direct boundary belong to $N(S)$.
\item Dual string operators that enclose a region that only contains direct boundaries belong to $S$.
\end{enumerate}
The second property implies that two dual string operators that together enclose such a region are equivalent up to stabilizers. Exchanging ``direct'' and ``dual'' we recover the defining properties of dual (rough) boundaries. All this is illustrated in Fig.~\ref{fig:rough}.

In Fig.~\ref{fig:planar_toric_code} we consider a somewhat more complicated geometry: four boundaries are combined to produce a planar toric code encoding a single qubit. Although this lattice can be obtained by removing two faces and a vertex in a sphere, there is no need to think this way: we only have to apply the rules above to understand the code in terms of string operators. Planar toric codes form a family of local $[[2d(d-1)+1,1,d]]$ codes.

\subsection{Error Correction}\label{sec:error_correction}

Our next goal is the analysis of the correction of Pauli errors $E$ in a surface code. A Pauli error $E$ can be written, up to unimportant phases, as $E=X_{c}Z_{\dual c}$. The first step in error correction is the measurement of stabilizer generators, in this case vertex and face operators. The resulting syndrome is dictated by \eqref{XZ_ops_boundary}: Vertex and face operators with eigenvalue $-1$ form the boundaries of $c$ and $\dual c$, respectively. According to the syndrome, any error $E\prima=X_{d}Z_{\dual d}$ with $(\partial d, \dual\partial {\dual d} )=( \partial c,\dual\partial {\dual c})$ may have happened. After choosing an applying such an $E\prima$, error correction will be successful if and only if $E\prima E\in S$. Since $E\prima E\propto X_{c\prima+d\prima}Z_{\dual c+\dual d}$, it follows that error correction succeeds if and only if $({c\prima+d\prima},{\dual c+\dual d})\in B_1\times \dual B_1$. That is, when errors and corrections belong to the same homology class, $(\overline c,\overline{\dual c})=(\overline d,\overline{\dual d})$. It is in this sense that in surface codes error correction must be done only up to homology, an advantage that has its origin in the fact that these are highly degenerate codes.

What is the best strategy to choose $E\prima$? Assume an error model in which Pauli errors $E=X_{c}Z_{\dual c}$ follow a probability distribution $\sset{p_{c,\dual c}}$. Rather than individual error probabilities, we are interested in the probability for the whole set of errors with the same effect in the code, namely
\begin{equation}\label{prob_class}
\prob{\overline c,\overline {\dual c}}:=\sum_{b \in B_1}\sum_{\dual b\in \dual B_1} \,p_{c+b,\dual c+\dual b}.
\end{equation}
The probability to obtain a given syndrome $(\partial c,\dual \partial \dual c)$ is
\begin{equation}\label{prob_synd}
\prob{\partial c,\dual\partial{\dual c}}:=\sum_{\dual z \in H_1}\sum_{\overline {\dual z}\in \dual H_1} \,\prob{\overline c+\overline z,\overline{\dual c}+\overline{\dual z}}.
\end{equation}
Given a particular syndrome $(\partial c, \dual\partial\dual c)$, the optimal strategy is to choose an $E\prima$ from the class of errors $(\overline d,\overline{\dual d})$ with highest conditional probability among those with  $(\partial d, \dual \partial \dual d)=(\partial c, \dual\partial {\dual c})$. The success probability is then
\begin{equation}\label{prob_cond}
p_\mathrm{max}(\partial c,\dual \partial \dual c) :=  \max_{(\overline z,\overline{\dual z})\in H_1\times \dual H_1} \frac{\prob{\overline c+\overline z,\overline{\dual c}+\overline{\dual z}}}{\prob{\partial c,\dual\partial\dual c}}.
\end{equation}
The overall success probability for this optimal strategy is recovered by weighting each syndrome with its probability, obtaining
\begin{multline}\label{prob_succ}
p_\mathrm{succ}:= \sum_{\partial c,\dual \partial \dual c} \prob{\partial c,\dual\partial \dual c} p_\mathrm{max} (\partial c,\dual\partial \dual c)=\\=\sum_{\partial c,\dual \partial \dual c} \max_{(\overline z,\overline{\dual z})\in H_1\times \dual H_1} \prob{\overline c+\overline z,\overline{\dual c}+\overline{\dual z}}.
\end{multline}
A very remarkable property of surface codes is that, in the limit of large lattices, $p_\mathrm{succ}\rightarrow 1$ when the noise is below a critical threshold. This will be the subject of next section.

In practice, computing which class has the maximal probability for a given syndrome might be costly, but there is an alternative approach. Notice that we can treat $X$ and $Z$ errors separately, ignoring any possible correlations. Then the problem reduces to choosing a 1-chain among those with a given boundary. When errors on physical qubits are independent, a good guess is to choose the chain $c$ with minimal weight, a problem that can be solved on polynomial time in the size of the lattice with the so-called perfect matching algorithm \cite{Dennis:2002:4452}. Since it is only an approximation, this technique provides a suboptimal critical threshold.

\subsection{Error Threshold: Random Bond Ising Model}\label{sec:threshold}

There exists a useful connection between the error correction threshold of surface codes and a phase transition in a 2D random bond Ising model. This connection appears when we separate, as above, the correction of bit-flip and phase-flip errors, ignoring any correlations. To fix ideas, we will study bit-flip erros, but phase flip errors have an analogous treatment. We also fix the geometry to that of toric codes, of variable size. We assume an error model where bit-flip errors occur independently at each qubit with probability $p$. Since bit-flip errors are represented by 1-chains $c$ as above, we have a probability distribution $\sset {p_c}$ with
\begin{equation}\label{pc}
p_c:=(1-p)^{E-|c|}p^{|c|}.
\end{equation}
Here $|c|$ denotes the number of edges of $c$ and $E$ the total number of edges. 

\subsubsection{Random Bond Ising Model}

Our first step is to define a family of classical Hamiltonian spin models. Classical spins are $s_i=\pm 1$ variables, with $i$ a label, and we attach one of them to each face. Alternatively, spins live at the vertices of the dual lattice, so that they are connected by edges of the dual lattice. As it is customary, we denote neighboring pairs of spins as $\pij$. We are interested in the family of classical Hamiltonians of the form
\begin{equation}\label{Hamiltonian_p}
H_\tau(s):= - \sum_{\langle ij \rangle}\, \tau_{ij} \,s_is_j,
\end{equation}
where the $\tau_{ij}=\pm 1$ are parameters of the Hamiltonian that define the ferromagnetic ($\tau_{ij}=1$) or antiferromagnetic ($\tau_{ij}=-1$) nature of the interactions. We have thus a family of Ising models with arbitrary interaction signs. These models exhibit a $\Z_2$ global symmetry, since flipping all spins at once does not change the energy. The equilibrium statistics of the system are described by the partition functions, which are
\begin{equation}\label{partition}
Z(\beta,\tau) = \sum_{s} e^{-\beta H_{\tau}(s)}.
\end{equation}
Here $\beta=1/T$ is the inverse temperature.

\begin{figure*}
\center
 \includegraphics[width=5cm]{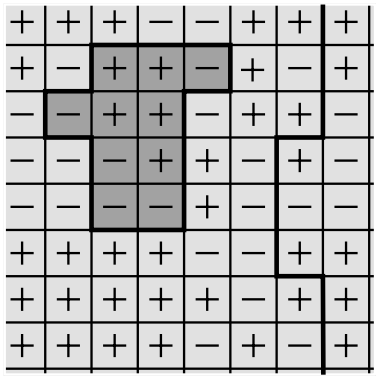}
\caption{The Ising model for a toric code. Classical $\pm 1$ spins live at plaquettes. Neighboring spins can interact ferro- or antiferromagnetically. If we switch all spins in the shaded region and at the same time also switch the sign of the interactions along its boundary, the energy is unchanged. But if we switch the sign along a nontrivial cycle like the one depicted by the bold line, this cannot be absorbed through spin switching.}
\label{fig:statistical}
\end{figure*}

Notice that the $\tau=\sset{\tau_{ij}}$ are 1-chains in disguise: since edges are labeled by pairs $\pij$, we can label the coefficients of 1-forms with such pairs, $c=\sset{c_{ij}}$ and then set $(\tau_c)_{ij}=(-1)^{c_{ij}}$. Similarly, we can attach to each $b\in B_1$ a spin configuration $s_b$ by choosing a 2-chain $d$ with $\partial d= b$ and setting $(s_b)_i:= (-1)^{d_i}$. Notice that here we are labeling 2-chain coefficients with spin labels, which is fine because spins live at faces of the direct lattice. Also, let as define a product on spin configurations:  $s\primas=s\prima \cdot s$ stands for $s\primas_i =s\prima_i s_i$.
We can now express in a simple way a crucial property of the model, illustrated in Fig.~\ref{fig:statistical}:
\begin{equation}\label{H_property}
H_{\tau_{c+b}}(s)=H_{\tau_c}(s\cdot s_b).
\end{equation}
Thus, homologically equivalent interaction configurations give equivalent systems, up to a transformation on the spin variables. On the other hand, if we change the sign of the interactions along a nontrivial loop, as the one of Fig.~\ref{fig:statistical}, this cannot be absorbed by a change of variables. For example, in a ferromagnetic system changing interactions along such a loop creates frustration: the new ground states gain an energy proportional to the length of the minimal loop with the same homology. This suggests introducing the notion of domain wall free energy for a given $\overline z\in H_1$, $\overline z\neq 0$:
\begin{equation}\label{free_energy}
\Delta_{\overline z} (\beta_p, \tau_c):=  F (\beta_p, \tau_{c+z}) -  F(\beta_p,\tau_c),
\end{equation}
where $F(\beta, \tau) = - T\log Z(\beta, \tau)$ is the free energy of a given interaction configuration $\tau$.

Rather than individual systems with Hamiltonian $H_\tau$, we are interested in the random bond Ising model, a statistical model obtained by making the parameter $\tau$ a quenched random variable. That is, $\tau$ is random but not subject to thermal fluctuations. We choose a probability distribution $\sset{p_\tau}$ such that each $\tau_{ij}$ has an independent probability $p$ of being antiferromagnetic. This will allow us later to connect with error correction, since $p_{\tau_c} = p_c$.

\begin{figure*}
\center
\psfrag{p}{$p$}
\psfrag{T}{$T$}
\psfrag{pc}{$p_\mathrm{crit}$}
\psfrag{Tc}{$T_\mathrm{crit}$}
\psfrag{p0}{$p_0$}
 \includegraphics[width=7cm]{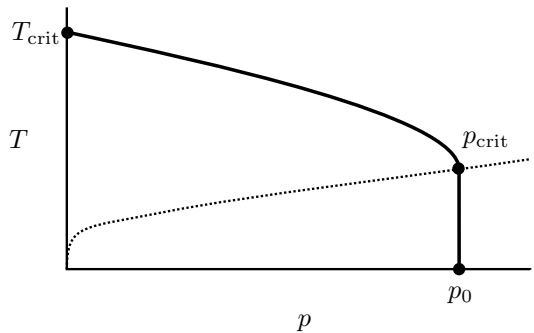}
\caption{Phase diagram of the random bond Ising model. $p$ is the probability of antiferromagnetic bonds and $T$ the temperature. The curve separates the low $p$, low $T$ ordered phase from the disordered phase. The Nishimori line is shown as a dotted line. The critical probability $p_\mathrm{crit}$ along the Nishimori line gives the error threshold for error correction.}
\label{fig:pT}
\end{figure*}

In thermal equilibrium the model has two parameters, the temperature $T$ and the probability $p$. For $p=0$, where we recover the standard Ising model, the system exhibits an order-disorder phase transition at a critical temperature $T_\mathrm{crit}$. For low temperatures the system is ordered, as it spontaneously breaks the global $\Z_2$ symmetry, and for high temperatures it is disordered. Order can also be destroyed at $T=0$ by increasing the disorder till we reach a critical value $p=p_0$. More generally, we can distinguish two regions in the pT plane, an ordered one at low $T$ and $p$, and a disordered one, as sketched in Fig.~\ref{fig:pT}. For the connection with error correction we will only be interested in the Nishimori line  \cite{Nishimori:1981:1169}, defined by
\begin{equation}\label{nishimori_line}
e^{-2\beta} = \frac p {1-p}.
\end{equation}
As we will see, the critical probability for error correction to be possible is given by the critical probability $p_\mathrm{crit}$ along the Nishimori line, see Fig.~\ref{fig:pT}. A witness of the ordering is the domain wall free energy
\begin{equation}\label{free_energy_p}
[\Delta_{\overline z}(\beta,\tau)]_{p}:= \sum_{\tau} p_\tau \,\Delta_{\overline z}(\beta,\tau),
\end{equation}
suitably averaged over quenched variables. This quantity diverges with the system size in the ordered phase and attains some finite limit in the disordered one.

\subsubsection{Mapping and Error Threshold}

In order to express the homology class probabilities $\prob{\overline c}$ in terms of the partition function \eqref{partition}, we first observe that
\begin{equation}\label{pc_H}
p_c = {(2\cosh \beta_p)^{-E}} {e^{-\beta_p H_{\tau_c}(s_0)}},
\end{equation}
with $\beta_p:= \log ((1-p)/p)/2$ the inverse temperature in the Nishimori line \eqref{nishimori_line} for a given $p$.
Using \eqref{H_property} we get the desired result:
\begin{equation}\label{}
\prob{\overline c}= 2^{-1} (2\cosh \beta_p)^{-E}\,Z(\beta_p,\tau_c),
\end{equation}
where the extra factor of 2 comes from the constraint \eqref{surface_code_stabilizers_constraint} or, equivalently, the global symmetry of the Ising model under spin inversion. We have then
\begin{equation}\label{}
\frac {\prob{\overline c+\overline z}}{ \prob{\overline c}}= e^{-\beta\Delta_{\overline z}(\beta_p,p)},
\end{equation}
which shows that the homology class $\overline c$ is much more probable than the other candidates when the domain wall energy is big. For the average success probability $p_\mathrm{succ}$ in 	\eqref{prob_succ} to be close to one, those syndromes that are most probable must be such that one class dominates, which implies a big average domain wall energy \eqref{free_energy_p}. Indeed, it can be shown that if $p_\mathrm{succ}\rightarrow 1$ then $[\Delta_z(\beta,\tau)]_{p}$ diverges \cite{bombin:2010:032301}, which establishes the connection between the error correcting threshold and the critical probability along the Nishimori line, which is
\begin{equation}\label{critical}
p_\mathrm{crit}\simeq 0.11.
\end{equation}

\section{Color Codes}

Surface codes are not very rich in terms of the gates that they allow to implement trough transversal operations. Since they are CSS codes, they allow the transversal implementation of CNot gates. And, of course, we can implement $\overline X$ and $\overline Z$ operators transversally. But that is all there is to it.

To go beyond these gates we need to consider a different class of topological codes: color codes. As we will see, this class of codes includes a family of planar codes that allow the transversal implementation of the whole Clifford group of gates.

\subsection{Lattice and Stabilizer Group}

\begin{figure*}
\center
\psfrag{r}{\normalsize r}
\psfrag{g}{\normalsize g}
\psfrag{b}{\normalsize b}
\psfrag{f}{\normalsize $f$}
 \includegraphics[width=7cm]{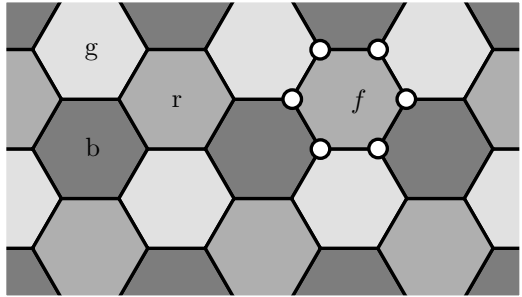}
\caption{Color codes are defined on 3-valent lattices with 3-colorable faces, like the honeycomb. We label faces as red, green and blue (r,g,b), as it is customary, and distinguish them with 3 tones of grey. Qubits are places on vertices. There are two generators of the stabilizer per face, $X_f$ and $Z_f$, with the support shown.}
\label{fig:color_code_honeycomb}
\end{figure*}

Surface codes can be built out of any lattice embedded in a closed manifold. In the case of color codes, we need to consider a particular kind of lattices: those that are 3-valent and have 3-colorable faces. That is, the lattice must be such that
\begin{enumerate}
\item three edges meet at each vertex and
\item it is possible to assign one of three labels to each face in such a way that neighboring faces have different labels.
\end{enumerate}
It is customary to choose as labels the colors red, green and blue: r,g,b. The most basic example of such a lattice is the honeycomb lattice, which can be embedded in a torus, see Fig.~\ref{fig:color_code_honeycomb}. Notice that in a color code lattice we can also attach a color to edges: red edges are those that do not form part of red faces, and similarly for green and blue edges.

In order to construct a color code from such a lattice the first step is to attach a qubit to each vertex. Next we need the stabilizer generators, of which there are two per face $f$:
\begin{equation}\label{color_code_stabilizers}
X_f:= \prod_{v\in f}\,X_v, \qquad\qquad Z_f:= \prod_{v\in f}\,Z_v,
\end{equation}
where $X_v,Z_v$ are the $X,Z$ Pauli operators on the qubit at the vertex $v$ and $v\in f$ is a symbolic notation to denote that $v$ is part of the face $f$.  Notice that, as surface codes, color codes are CSS codes with local generators. The  ``plaquette'' operators \eqref{color_code_stabilizers} are shown in Fig.~\ref{fig:color_code_honeycomb}.

As in \eqref{surface_code_stabilizers_constraint}, the stabilizer generators \eqref{color_code_stabilizers} are not independent. They are subject to four constraints. In order to write them, let us denote by $\fr$, $\fg$ and $\fb$ the sets of red, green and blue faces, respectively. It is not difficult to check that the constraints are
\begin{alignat}{2}\label{color_code_stabilizers_constraint}
\prod_{f \in\fr} X_f&=\prod_{f \in\fg} X_f&=\prod_{f \in\fb} X_f,\nonumber\\
\prod_{f \in\fr} Z_f&=\prod_{f \in\fg} Z_f&=\prod_{f \in\fb} Z_f.
\end{alignat}
Thus, the number of independent stabilizer generators is
\begin{equation}\label{color_indep}
g=2(|\fr|+|\fg|+|\fb|)-4.
\end{equation}
This allows us to immediately compute the number of encoded qubits. Namely, since there are $2E$ physical qubits
\begin{equation}\label{color_k1}
k=n-g=2(E-|\fr|-|\fg|-|\fb|+2).
\end{equation}
However, to express $k$ in terms of topological invariants we need a geometrical construction, which is our next topic.

\subsection{Shrunk Lattices}

\begin{figure*}
\center
 \includegraphics[width=7cm]{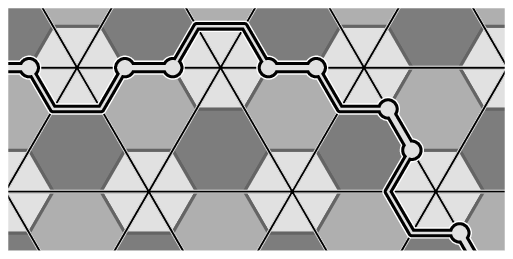}
 \caption{Any shrunk lattice of a honeycomb lattice is triangular. Here we show the green shrunk lattice and a string $\gamma$. The qubits in the support of $X_\gamma^\mathrm{g}$ and $Z_\gamma^\mathrm{g}$ are marked with circles along the string. They come in pairs, with each pair related to an edge of the green shrunk lattice.}
\label{fig:shrunk_lattice}
\end{figure*}

Out of a color code lattice we want to build three other ``shrunk'' lattices, labeled by the color of the faces that are actually shrunk. Let us focus on the red shrunk lattice; the green and blue are analogous. The vertex of the new lattice correspond to red faces, which are in this sense shrunk to points. Edges come from those edges that connect red faces, that is, red edges. As for faces, there is one for each green and blue face of the original lattice. The construction is demonstrated in Fig.~\ref{fig:shrunk_lattice}.

Now, let $V^\mathrm{r},E^\mathrm{r},F^\mathrm{r}$ denote the number of vertices, edges and faces of the red shrunk lattice. We get using (\ref{Euler}, \ref{Euler_genus}, \ref{color_k1}) that the number of encoded qubits is
\begin{equation}\label{}
k=2(E^\mathrm{r}-V^\mathrm{r}-F^\mathrm{r}+2)=2(2-\chi)=4g.
\end{equation}
That is, encoded qubits have a topological origin! Notice that the number of encoded qubits doubles that of surface codes. The origin of this doubling will become clear when we explore string operators.

\subsection{String Operators}

\begin{figure*}
\center
\psfrag{a}{\normalsize $a$}
\psfrag{b}{\normalsize $b$}
\psfrag{c}{\normalsize $c$}
\psfrag{d}{\normalsize $d$}
\psfrag{e}{\normalsize $e$}
 \includegraphics[width=9cm]{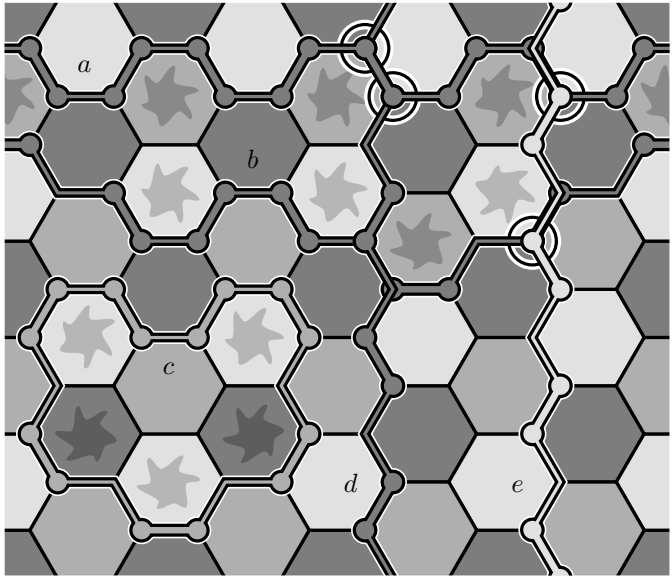}
\caption{A honeycomb color code in a torus, with five closed strings on display. For each string the support of the string operators for a given color are shown: blue for $a,b,d$, red for $c$ and green for $e$. Those qubits at which two string operators share support are marked with big circles. Strings $a$ and $b$ are homologous and thus, for $\sigma=X,Z$, we have $\sigma_a^\mathrm{b}=A\sigma_b^\mathrm{b}$ with $A\in S$ the product of the face operators $\sigma_f$ marked between $a$ and $b$. String $c$ is homologically trivial and thus $\sigma_c^\mathrm{r}\in S$: it is obtained as the product of face operators $\sigma_f$ marked in the region enclosed by $c$. Strings $d$ and $e$ are homologous, but due to the different colors $\sigma_d^\mathrm{b}\sigma_e^\mathrm{g}\nin S$. Strings $a$ and $d$ cross, but because the color is the same we get $[X_a^\mathrm{b},Z_d^\mathrm{b}]=0$, and similarly for $b$ and $d$. Strings $a$ and $e$ cross, and due to the different coloring we have $\sset{X_a^\mathrm{b},Z_e^\mathrm{g}}=0$, and similarly for strings $b$ and $e$. }
\label{fig:color_code_strings}
\end{figure*}

Given a loop or closed string $\gamma$ in a color code lattice, we can construct out of it six different string operators: $X_\gamma^c$ and $Z_\gamma^c$,with $c= \mathrm{r}, \mathrm{g}, \mathrm{b}$ a color.
They take the form
\begin{equation}\label{color_code_strings}
X_\gamma^c := \prod_{v\in V_c^{\gamma}} X_v,\qquad\qquad Z_\gamma^c := \prod_{v\in V_c^{\gamma}} Z_v,
\end{equation}
where the set $V^c_{\gamma}$ contains those vertices that belong to a $c$-colored edge of $\gamma$. Indeed, the support of a string operator is best understood in terms of the shrunk lattice, as explained in Fig.~\ref{fig:color_code_strings}.

We next list several properties of string operators that are easy to check. String operators obtained from closed strings belong to the normalizer $N(S)$. For a given closed string $\gamma$ there are only four independent string operators because
\begin{equation}\label{color_code_string_constraints}
X_\gamma^r X_\gamma^g X_\gamma^b = 1,\qquad\qquad Z_\gamma^r Z_\gamma^g Z_\gamma^b = 1.
\end{equation}
In this sense, there are only two independent colors, so that it suffices to consider, say, red and green strings. If two strings $\gamma$ and $\gamma\prima$ cross an even number of times their operators commute. If they cross an odd number of times, we have $[X_{\gamma}^c,Z_\gamma^c]=0$ and $\sset{X_{\gamma}^c,Z_\gamma^{c\prima}}=0$ for $c\neq c\prima$, see Fig.~\ref{fig:color_code_strings}.

A question without such an immediate answer is: when does a string operator belong to the stabilizer and when are two string operators equivalent as logical operators in $N(S)/S$? As in surface codes, the answer lies in homology: the trick is to think in terms of the shrunk lattice. For example, consider $X$-type red string operators. We can regard any $\gamma$ as a loop in the red shrunk lattice. If $\gamma$ and $\gamma\prima$ are homologically equivalent,
they enclose a region in the shrunk lattice. This region corresponds to a set of green and blue faces in the original lattice. As one can easily check, we have then  $X_\gamma^r=sX_{\gamma\prima}^r$ with $s\in S$ the product of the corresponding $X$-type face operators. It follows that boundary strings produce elements of $S$ and that homologically equivalent strings produce, for each type of operator, equivalent operators up to stabilizers. All this is illustrated in Fig.~\ref{fig:color_code_strings}.

\begin{figure*}
\center
\psfrag{X1Z2}{\normalsize $\overline X_1, \overline Z_2$}
\psfrag{X3Z4}{\normalsize $\overline X_3, \overline Z_4$}
\psfrag{X2Z1}{\normalsize $\overline X_2, \overline Z_1$}
\psfrag{X4Z3}{\normalsize $\overline X_4, \overline Z_3$}
 \includegraphics[width=9cm]{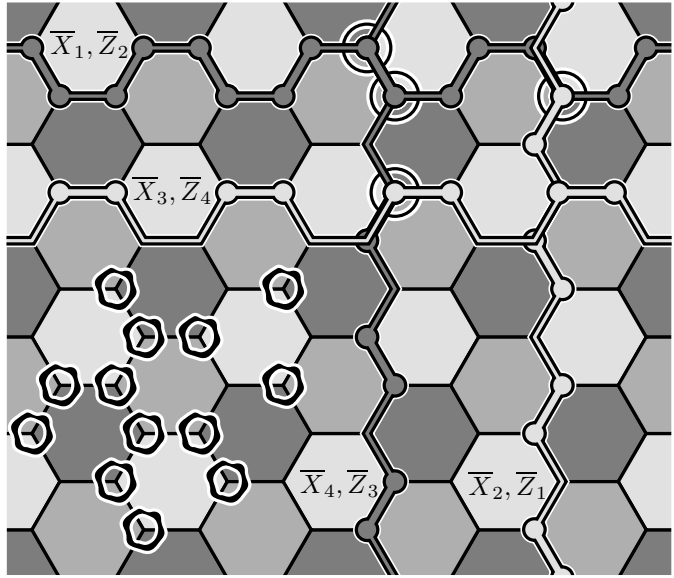}
\caption{A [[96,4,8]] color code in a torus. Logical operators $\overline X_i, \overline Z_i$ take the form of four blue string operators and four green string operators. The marked qubits form the support of a local operator. If such an operator belongs to $N(S)$, then it belongs to $S$ because it clearly commutes with logical operators.}
\label{fig:torus_color_strings}
\end{figure*}

We are now ready to choose logical operators for a given surface. Indeed, the task is almost the same as in surface codes, but now we have to take color into account. In particular, strings of two colors are needed, which is at the origin of the doubling of encoded qubits with respect to surface codes. As an example, we show a choice of logical operators for a torus in Fig.\ref{fig:torus_color_strings}. We adopt the convention that $\overline X_i$ and $\overline Z_i$ logical operators are obtained respectively from $X$-type and $Z$-type string operators.

The fact that logical operators can be chosen to be string operators has an important consequence that is not specific of color codes but common to 2D topological codes. If a region $R$ is such that we can choose a set of logical operators with support out of it, any operator with support in $R$ that belongs to $N(S)$ must belong to $S$. Since strings can be deformed, this is in particular true for any ``local'' region, such as the one in Fig.~\ref{fig:torus_color_strings}. In the case of regular lattices and due to the locality of stabilizer generators, this implies that the code distance will grow with the lattice. Thus, color codes are indeed local codes.

\subsection{String-Nets}

We could be tempted to believe that the distance in a color code is given by the smallest weight among string operators of nontrivial homology. This holds in all the examples that we shall present, but in general it is not true. The reason is that we can combine strings to form nets, resulting in smaller weights.

\begin{figure*}
\center
\includegraphics[width=8cm]{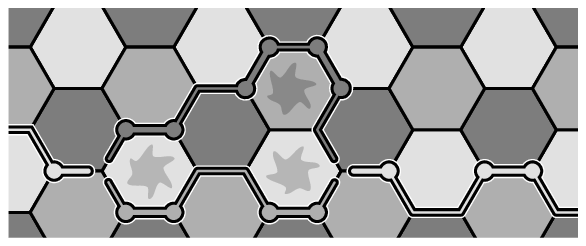}
\caption{A string-net operator. It can be transformed into a green string operator by taking the product with the face operators from the marked faces.}
\label{fig:string_net}
\end{figure*}

In order to understand what these string-nets are, take any green string and consider deforming part of it not by taking the product with the blue and red face operators in an adjacent region, as we should, but with the red and green face operators, as if it were a blue string. The result, as Fig.~\ref{fig:string_net} shows, is not a string any more, but a net of three strings. The deformation has created a piece of blue string, leaving a red string where the piece of green string to be deformed was. The moral is that strings can have branching points and form string-nets. At each branching point three strings, one of each color, must meet. Although string-nets are not necessary to construct logical operators in closed surfaces, we will see how they can become essential in the presence of boundaries.

\subsection{Boundaries}

As in surface codes, in color codes we can obtain boundaries by erasing ``big'' faces from the lattice. There are thus three kinds of boundaries, one per color. To make the idea clear, we consider blue boundaries. These are obtained by erasing a blue face, so that blue strings can have endpoints at the boundaries, but not green or red ones. The properties of boundaries in color codes are analogous to those in section \ref{sec:boundaries}:
\begin{enumerate}
\item Blue string operators that have their two endpoints on a blue boundary belong to $N(S)$.
\item Blue string operators that enclose a region that only contains blue boundaries belong to $S$.
\end{enumerate}
It is worth noting that these three kinds of boundaries are not exhaustive. For example, it is possible to have boundaries where only $X$-type string operators can have endpoints. But we will not need this more general cases.

From the constraints \eqref{color_code_stabilizers_constraint} we can infer the number of encoded qubits in geometries with boundaries: For each new face that we erase, we add two qubits, unless we have only erased a single face of a different color previously. In this latter case we add no qubits. For example, if in a sphere we remove one face of each color the resulting color code encodes two qubits. We will next discuss a closely related geometry that encodes a single qubit.

\subsection{Transversal Clifford Group}

\begin{figure*}
\center
\psfrag{X}{\normalsize $\overline X$}
\psfrag{Z}{\normalsize $\overline Z$}
 \includegraphics[width=10.5cm]{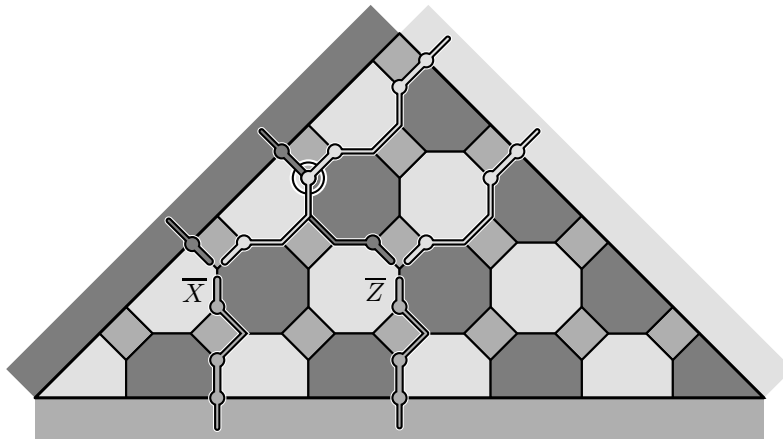}
\caption{A $[[73,1,9]]$ triangular color code, based on a 4-8-8 lattice. The bottom boundary is red, the right one green and the left one blue. There is only one class of nontrivial string-nets, call it $\mu$. That $X_\mu$ and $Z_\mu$ anticommute can be understood topologically by considering the string-net and a deformation of it, as in this figure, and noting that they cross once at a point where they have different colors.}
\label{fig:triangular_code}
\end{figure*}

We can finally introduce the family of color codes that allows the transversal implementation of the whole Clifford group of gates: triangular codes. From a topological perspective, these are planar codes with the geometry of a triangle in which each side is a boundary of a different color, as depicted in Fig.\eqref{fig:triangular_code}. How many qubits are encoded with such a topology? A bit of experimentation can convince one that there is only one nontrivial class of string-nets, shown in the figure. There is something very special about this string-net. Denote it by $\mu$. Then we have $\sset {X_\mu,Z_\mu}=0$, so that the encoded Pauli operators $\overline X=X_\mu$ and $\overline Z=Z_\mu$ can be chosen to have the same geometry! 

Notice that any color code is invariant under the action of a transversal Hadamard gate $\widehat W:=W^{\otimes V}$, where $V$ is the number of vertices/qubits, because $\widehat W X_f \widehat W =  Z_f$ and $\widehat W Z_f \widehat W = X_f$. For the triangular geometry we have in addition $\widehat W \overline X \widehat W=\overline Z$ and $\widehat W \overline Z \widehat W=\overline X$, simply because geometrically $\overline X$ and $\overline Z$ are the same.

Since CNot gates are automatically transversal in a CSS code, all we need is to find a way to implement the phase gate $P$ transversally. The obvious guess is to use $\widehat P := P^{\otimes V}$ but, does it leave the code invariant? In general no, because $\widehat P Z_f \widehat P=Z_f$ but $\widehat P X_f \widehat P=(-1)^{t/2}X_fZ_f$, with $t$ the number of vertices of the face $f$. All we have to do then is to find lattices where all faces have a number of edges that is a multiple of four. This is indeed possible using the so-called 4-8-8 lattice, as in Fig.~\ref{fig:triangular_code}. As for the effect of $\widehat P$, it gives either an encoded $P$ or $-P$, because $\overline X$ always has support on an odd number of qubits (otherwise, it could not have the same support as $\overline Z$). As a result, we have obtained a family of 2D local codes that allow the transversal implementation of any Clifford gate.

\subsection{Homology}

In the case of surface codes, we started by giving a homological definition and from that we obtained a description in terms of a stabilizer group. For color codes the definition has been in terms of the stabilizer, but we can now obtain from it a homological description by undoing our steps for surface codes.

\begin{figure*}
\center
\psfrag{d1}{\normalsize $\partial_1$}
\psfrag{d2}{\normalsize $\partial_2$}
 \includegraphics[width=10.5cm]{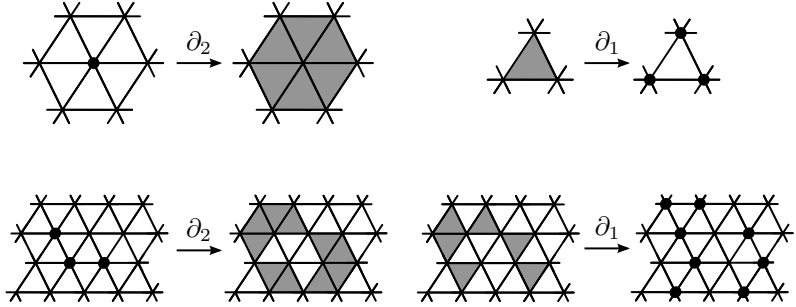}
 \caption{The action of ``color'' boundary operators. $\partial_2$ maps a set of vertices to the set of triangles to which an odd number of vertices belong. $\partial_1$ maps a set of triangles to to the set of vertices where an odd number of these triangles meet.}
\label{fig:boundary_color}
\end{figure*}

Before we do this, it is useful to change our picture of color codes by switching to the dual lattice. The dual of a color code lattice has triangular faces and three colorable vertices. For example, the honeycomb lattice has as its dual the triangular lattice of Fig.~\ref{fig:boundary_color}. Notice that in the dual picture qubits are attached to triangles and stabilizer generators to vertices.

In our search for the ``color homology'', we start observing that, since qubits are attached to triangles, 1-chains should be composed of triangles. We have thus a group $C_1^\triangle$ with elements that are formal sums of triangles. As for 0-chains and 2-chains, they must be composed of the geometrical objects attached to $Z$ and $X$ stabilizer generators, respectively. For color codes they are the same: we take the elements of $C_0^\triangle=C_2^\triangle$ to be formal sums of vertices.

The next step is to build the boundary morphisms $\partial$. This is dictated by the geometry of stabilizer generators. The morphisms $\funcion{\partial_2^\triangle}{C_2^\triangle}{C_1^\triangle}$ and $\funcion{\partial_1^\triangle}{C_1^\triangle}{C_0^\triangle}$ are dual to each other:
\begin{equation}\label{triangle_boundaries}
\partial_2^\triangle v = \sum_{f|v\in f} f,\qquad\qquad \partial_1^\triangle f = \sum_{v\in f} v.
\end{equation}
where $v$ stands for a vertex and $f$ for a triangular face. The action of the boundary operators is illustrated in Fig.~\ref{fig:boundary_color}. Boundary morphisms give rise to a group of cycles $Z_1^\triangle$ and a group of boundaries $B_1^\triangle$, with $B_1^\triangle\subset Z_1^\triangle$. In a closed surface the resulting homology group must be
\begin{equation}\label{triangle_homology}
H_1^\triangle:= \frac{Z_1^\triangle}{B_1^\triangle}\simeq H_1\times H_1,
\end{equation}
because we know that there are two independent ``homology structures'', one per independent color.

Once we have a homological language for color codes, we can immediately apply all the results on error correction of section \ref{sec:error_correction}. Color codes also attain in the limit of large lattices $p_\mathrm{succ}\rightarrow 1$ when the noise is below a critical threshold. The main difference with error correction in surface codes is algorithmic, at least under the simplifying approach that led to length minimization. Here this approach leds to the problem of finding a triangle chain of minimum weight for a given boundary, which cannot be solved using the perfect matching algorithm.

\subsection{Error Threshold: Random 3-Body Ising Model}

\begin{figure*}
\center
 \includegraphics[width=9cm]{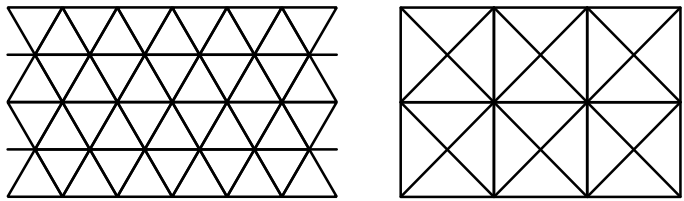}
\caption{Triangular lattice (left) and Union Jack lattice (right). These are dual of the honeycomb and 4-8-8 lattices, respectively.}
\label{fig:triangular_lattices}
\end{figure*}

As in the case of surface codes, the error correction threshold of color codes is connected to a phase transition in a 2D statistical model: a random 3-body Ising model. Given the similarities of the mappings, we will only discuss those aspects that are different with respect to surface codes, and the assumptions are the same. We consider two geometries, the honeycomb lattice and the 4-8-8 lattice that allows the transversal implementation of $P$. Or rather, the duals of these lattices, the triangular lattice and the so-called Union Jack lattice, shown in Fig.~\ref{fig:triangular_lattices}.

A question that comes to mind immediately is: do the transversality properties have a cost in terms of the error threshold? Surprisingly, the answer turns out to be negative.

\subsubsection{Random 3-Body Ising Model}

This time classical spins are attached to the vertices of the dual lattice, so that we can talk of red, green and blue spins. Each triangular face can be given as a triad of vertices $\tij$. The Hamiltonians take the form
\begin{equation}\label{}
H_\tau(s):= - \sum_\tij\, \tau_{ijk} \,s_is_js_k,
\end{equation}
where the $\tau_{ijk}=\pm 1$ are again parameters of the Hamiltonian that determine the sign of the 3-body interactions. Instead of the $\Z_2$ symmetry of the 2-body Ising models, these models exhibit a $\Z_2\times \Z_2$ global symmetry: for any color of choice, flipping all the spins of the two other colors leaves the energy unchanged. Equation \eqref{H_property} still holds, with the obvious definitions for $\tau_c$ and $s_b$, $c\in C_1^\triangle$, $b\in B_1^\triangle$. The rest of details of the model are analogous to those for toric codes, including the phase diagram.

\subsubsection{Mapping and Error Threshold}

The mapping works essentially as in toric codes. There is only a slight difference, that the factor due to global symmetry is now 4:
\begin{equation}\label{probability_partition_color}
\prob{\overline c}= 4^{-1} (2\cosh \beta_p)^{-F}\,Z(\beta_p,\tau_c),
\end{equation}
where $F$ is the number of triangles. As for the critical probability, for both lattices we have
\begin{equation}\label{}
p_\mathrm{crit}\simeq 0.11,
\end{equation}
the same as for toric codes!

\section{Conclusions}

Topological codes are naturally local, which makes them appealing for practical implementations with locality constraints. We have described two classes of topological codes, surface codes and color codes. The main difference between them is that color codes allow the transversal implementation of more gates, even of all the gates in the Clifford group. Although topological codes were initially described in closed surfaces, it is possible to construct planar versions by introducing carefully designed boundaries.

We have emphasized the role of homology in topological codes, which offers a unified picture of surface and color codes. The first homology group is obtained as a quotient $Z_1/B_1$, and the very essence of topological codes is the identification of this quotient with $N(S)/S$, the quotient that gives logical operators in stabilizer codes. That is, stabilizer elements correspond to boundary loops, normalizer elements to closed loops and logical operators to elements of the homology group.

From the point of view of error correction, topological codes exhibit two remarkable facts. One is that error correction must be done only up to homology, due to the high degeneracy of the codes. The other is the existence of an error threshold: for noise levels below this threshold, in the limit of large systems error correction is asymptotically perfect. For error models with uncorrelated bit-flip and phase flip errors, those for which the critical threshold is well understood, the critical error probability is $p_\mathrm{crit}\simeq 0.11$.

\section{History and Further Reading}

Topological codes started their history with the introduction by Kitaev of the toric code \cite{kitaev:1997:181, kitaev:2003:2}. It was soon realized that boundaries allow one to build planar codes \cite{bravyi:1998:9811052, freedman:2001:325}. The basic reference on surface codes is \cite{Dennis:2002:4452}. This work introduced, among other things, the concept of topological quantum memory: in the limit of large surface codes, there exists an error threshold below which quantum information can be preserved arbitrarily well. It also discusses higher dimensional toric codes and shows the connection between accuracy thresholds in error correction and phase transitions in statistical models, a subject developed in \cite{wang:2003:31, ohno:2004:462, takeda:2004:377} and other works. Another concept introduced in \cite{Dennis:2002:4452} is that of code deformation: the lattice defining the code can be progressively transformed locally, for example to encode information by ``growing'' the lattice as a crystal. It was later realized that this can be used to initialize, measure and perform gates on encoded qubits \cite{raussendorf:2007:199,bombin:2009:095302}, something closely related to the fault-tolerant one-way quantum computing scheme of \cite{raussendorf:2006:2242}. Recent examples of how research on toric codes continues more than a decade after their introduction are a study on loss errors \cite{stace:2009:200501} and a new algorithm for error correction based on renormalization \cite{duclos:2010:50504}.

2D color codes were introduced in \cite{bombin:2006:180501} and soon a generalization to 3D followed \cite{bombin:2007:160502}. The advantage of 3D color codes is that they allow the transversal implementation of the CNot gate, the $\pi/8$ phase gate and $X$, $Z$ measurements: a universal set of operations for quantum computing. The statistical models related to error correction in 2D color codes have been recently studied in \cite{katzgraber:2009:090501,katzgraber:2009:012319}.

Surface codes and color codes are not the end of the story. Other interesting topological codes might still be awaiting their discovery. Among recent developments we find topological subsystem codes \cite{bombin:2010:032301} and Majorana fermion codes \cite{bravyi:2010:083039}. New ways to exploit already known codes are also valuable. Twists, recently introduced in \cite{bombin:2010:30403}, exemplify this, as they offer a new tool for constructing planar topological codes with enhanced code deformation capabilities.

Topological codes give rise naturally to condensed matter systems in which the ground state corresponds to the encoded subspace \cite{kitaev:2003:2}. These \emph{topologically ordered }systems are stable against perturbations at $T=0$: local modifications of the Hamiltonian, if not too big, do not affect the physics \cite{bravyi:2010:10010344}. The effect of thermal noise depends on the spatial dimension of the system: in two dimensions topological order is destroyed \cite{alicki:2009: 065303}, but in four dimensions it can be resilient to noise \cite{alicki:2008:08110033}, giving rise to a self-correcting quantum memory. In six dimensions it is even possible to put together such \emph{self-correcting} capabilities and the nice transversality properties of 3D color codes \cite{bombin:2009:09075228} to obtain a self-correcting quantum computer.

The excitations exhibited by 2D topologically ordered systems are gapped and localized. This quasiparticles, called \emph{anyons}, have unusual statistics, neither bosonic nor fermionic. Indeed, they give rise to non-local, topological degrees of freedom that can be manipulated by moving the anyons around. This offers a new way to perform quantum computations: topological quantum computation \cite{kitaev:2003:2,freedman:2003:31}. A good introduction to this topic is \cite{Preskill:2004:caltech}. In higher dimensions excitations take the form of extended objects called branyons in \cite{bombin:2007:75103}.

\bibliography{mybib}{}
\bibliographystyle{ieeetr}
\end{document}